# Magnetic polaronic exciton in A-type 2D van der Waals bulk material CrSBr

*Xiaodong Shen[1], Jiajun Cao[1], Weizheng Liang[1]\*, Borong Cong[1], Bao Ke[1], Jialong Zhao[1], Bingsuo Zou[1]\**

1, *State Key Laboratory of Featured Metal Materials and Life-cycle Safety for Composite Structures, MOE Key Laboratory of New Processing Technology for Nonferrous Metals and Materials; School of Physical Science and Technology; School of Resources, Environment and Materials; Guangxi University, Nanning 530004, China;*

*\*zoubs@gxu.edu.cn ; orcid.org/0000-0003-4561-4711*

## Abstract

2D magnetic semiconductor CrSBr exhibits unique magneto-optical properties, yet its electronic structure and photophysical mechanisms remain unclear at high magnetic field and low temperature. Through comprehensive spectroscopic investigations, its charge-transfer band edge is identified at 500 nm. Below this band-edge, local excitonic magnetic polaronic states from $Cr^{3+}$ ions out of FM aggregates in layer and bilayer could be seen due to phonon-spin-exciton coupling, in which magnetic polaronic $PL_1$ emission occurs at 720 nm from single $Cr^{3+}$ d-d transition, a dark-state pair exciton occurs at 850 nm in 10 K magnetic field, and double-peak $PL_2$ emission at 920 nm out of $Cr^{3+}$ FM trimer in monolayer is seen; besides, the magnetic bi-polaronic $PL_3$ at 990 nm can be assigned to $Cr^{3+}$ tetramers between FM adjacent layers. In magnetic field perpendicular to the layer, direct competition between $PL_1$/dark-state excitons and $PL_2$/$PL_3$ excitonic states persist in different temperatures. This study sheds light on the complicated magneto-exciton interactions in the multi-body effect of CrSBr, beneficial for quantum modulation in layered magnetic semiconductors.

# Introduction

Two-dimensional (2D) van der Waals (vdW) materials have attracted significant research interest due to their unique exciton effects[1-3]. In these materials, exciton-neutral bosonic quasi-particles of electron-hole pairs bound by Coulomb interaction exhibit much higher binding energies than traditional semiconductors, leading to pronounced impacts on their optical and electrical properties[3-8]. CrSBr, a layered antiferromagnetic (AFM) semiconductor, represents a typical system for studying such effects. The crystal structure of CrSBr consists of vdW layers composed of two CrS layers terminated by Br atoms, stacking along the *c-axis* to form an orthogonal structure[9]. Each monolayer shows ferromagnetic (FM) ordering at 140 K, while in bulk form, these FM layers may couple antiferromagnetically below the Néel temperature ($T_N$ = 132 K) [10,11]. CrSBr shows biaxial magnetic anisotropy, with its easy magnetic axis along the *b-axis*, intermediate axis along the *a-axis*, and hard axis along the *c-axis*. It is a direct-gap semiconductor with an electronic gap of 1.5 eV and an excitonic gap of 1.34 eV (also exists in a single layer) [12-16]. Besides these, high-energy $PL_1$ and lower $PL_3$ excitons was found[17,18]. Although extensive studies have been conducted on Cr$^{3+}$-doped materials' optical properties[19-21], particularly their near-infrared emissions[22,23], the detailed spin-related electronic structure and emission mechanisms remain not well understood. This paper investigates the energy levels and optical processes in CrSBr through various spectroscopic techniques, including temperature-dependent photoluminescence (PL), femtosecond transient absorption (TA) spectroscopy, and magnetic field-dependent measurements. Our low-temperature (10 K) PL studies reveal different Cr$^{3+}$ ion aggregate states with distinct emission profiles and identify new dark-state excitons in magnetic field. These findings suggest more complicated microscopic interactions between spin-phonon-exciton in the layered structure than previously thought, providing insights into the interplay between $Cr^{3+}$ ions, modified d-d transitions and PL features.

## Results and discussion

CrSBr, a 2D vdW magnetic semiconductor belonging to the $P_{mmn}$ space group, features a layered structure and biaxial magnetic anisotropy (with easy *b-axis*, hard *c-axis*, and intermediate *a-axis*) [24,25]. High-quality bulk crystals were prepared via chemical vapor transport and confirmed by EDS with 1:1:1 atomic ratio. Three characteristic Raman peaks are observed at room temperature under 532 nm laser excitation: $A_g^1 = 113$ cm$^{-1}$, $A_g^2 = 242$ cm$^{-1}$, $A_g^3 = 339$ cm$^{-1}$ (Fig S1a), At 10 K under 488 nm laser excitation, a peak at 120 cm$^{-1}$ appears, indicating phonon zone folding due to long-range electron-phonon coupling. Under 488 nm laser excitation at room temperature, a weak high-energy $PL_1$ band appears at 720 nm. This band is enhanced relative to the 950 nm infrared emission when excited by 325 nm laser or 315 nm Xenon lamp (Fig S1b and Fig S9a). At low temperature (10 K), the infrared PL splits into a double-peak $PL_2$ band at 920 nm and a weak $PL_3$ band about 990 nm (Fig 1c and Fig 3c). The double-peak feature of $PL_2$ mainly arises from exciton-phonon coupling under 488 nm excitation. When exciting fewer layers (less than 5) with 488 nm laser, only $PL_1$ and $PL_2$ are observed (Fig S3c). In contrast, under 400 nm excitation, $PL_1$ almost disappears while $PL_3$ becomes stronger than $PL_2$. Polarization measurements using 488 nm linearly polarized laser excitation reveal that $PL_1$ and $PL_2$ polarization directions differ by 13.02° (Fig S14d). These observations demonstrate that $PL_1$, $PL_2$, and $PL_3$ originate from different exciton states out of spin-phonon-exciton interactions in CrSBr.

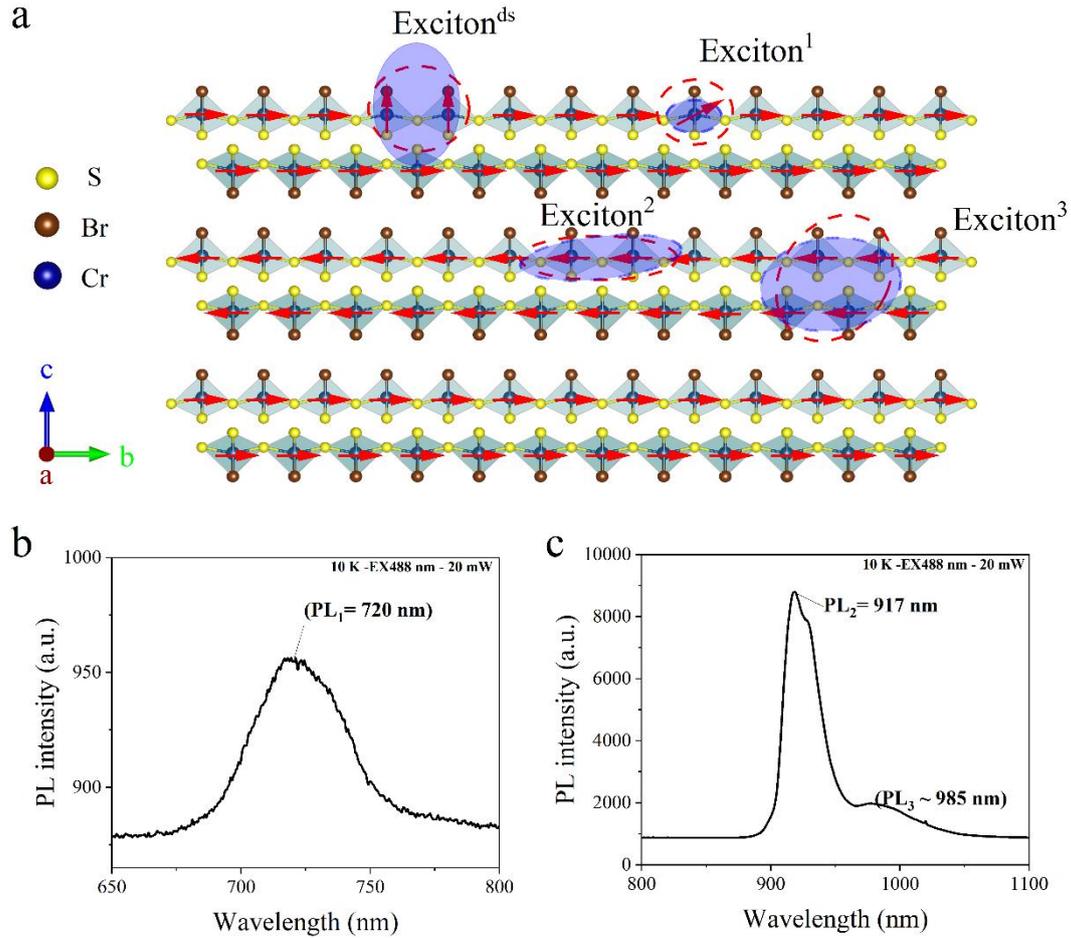

Figure 1 (a) Schematic diagram of multiple excitons in bulk CrSBr at 10 K, showing: exciton[1] (formed by a single $Cr^{3+}$ ion), exciton[2] (a trimeric exciton with three FM spins aligned along the *b*-axis), and exciton[3] (a tetrameric exciton in with FM double-layers). Dark-state exciton (exciton[ds]), dimeric excitons aligned along the *c-axis*. translucent blue regions represent electric dipoles, and red arrows indicate spin directions. (b) PL spectra measured under 488 nm laser excitation at 20 mW and 10 K showing: $PL_1$ at 720 nm from single $Cr^{3+}$ ion (exciton[1]), $PL_2$ at 917 nm from the $Cr^{3+}$ FM trimer (exciton[2]), and $PL_3$ at ~985 nm corresponding to the interlayer $Cr^{3+}$ FM tetramer(exciton[3]).

Various d-d level emissions and transitions are observed in bulk CrSBr. Room temperature UV-Vis absorption measurements reveal multiple characteristic d-d transitions: a main absorption peak at 813 nm (1.53 eV) with additional absorption bands at 300 nm, 472 nm, and 600 nm, corresponding to $^4A_2 \rightarrow {^4T_1}$, $^4A_2 \rightarrow {^4T_1}$, and $^4A_2 \rightarrow {^4T_2}$ transitions, respectively[2,26]. At 0 T and 10 K, $PL_1$ shows linear power dependence (Fig 2a-b, 1-20 mW), characteristic of a single $Cr^{3+}$ ion $^4A_2 \rightarrow {^4T_2}$ spin-allowed d-d transition in transition metal-doped semiconductors[27], where this linearity reflects direct excitation at low powers and the small oscillator strength of d-d transitions compensates for the state's long lifetime(Fig S9b), maintaining linearity even at high excitation powers[28]. $PL_2$ exhibits nonlinear power dependence with enhanced coherence and electron-phonon coupling (Fig 2c-d), characterizing localized magnetic polaronic exciton (EMP) [27,29] in a single layer, with rapid relaxation from $^4T_2$ states in multilayer structures, eventually merging with $PL_3$ at higher powers (Fig S4, 40 mW). Moreover, the $PL_3$ emission band diminishes at higher power because these local transitions can be decoupled due to carrier scattering[30]. This decoherence-related scattering can potentially be attributed to two mechanisms: an enhancement of rapid relaxation from the carrier localization out of $^4T_2$ states within the two-layer structure, consequently rendering the $PL_3$ peak disappearance, returning to the $PL_2$ states at high power excitation. PLE measurements reveal two primary bands at 438 nm (combined CT bands and $^4A_2 \rightarrow {}^4T_1$) and 569 nm ($^4A_2 \rightarrow {}^4T_2$), Below 132 K, which corresponds to the AFM phase transition temperature, a third band emerges at 733 nm (Fig S6a). notably, this $PLE_3$ (Corresponds with $PL_3$) excitation band around 733 nm almost disappears above $T_N$ due to forbidden transition rules in the paramagnetic phase (Fig S6a). Examination of PLE profiles at different emission wavelengths (Fig S6b, 915-992 nm) confirms that $PL_2$ and $PL_3$ originate from the same coupled d states, where we designate the second and third PLE bands as $^4T_2^{(1)}$ and $^4T_2^{(2)}$, respectively, with $^4T_2^{(2)}$ potentially arising from the absorption of FM $Cr^{3+}$ aggregates. The $PL_3$ band is exclusively observed in bulk materials and is essentially invisible in near-monolayer samples, with studies confirming its presence only in multilayer structures exceeding 15 layers[23,25]. Below 132 K, bulk CrSBr transforms into the AFM phase characterized by three major PLE peaks, and this $PL_3$ band vanishes above $T_N$ in the paramagnetic phase, suggesting its origin from interlayer excitons involving four $Cr^{3+}$ ions (two per layer) in the neighboring FM bilayer (Fig 1a). Therefore, this AFM phase are per-two layer to convert their spins in this compound.

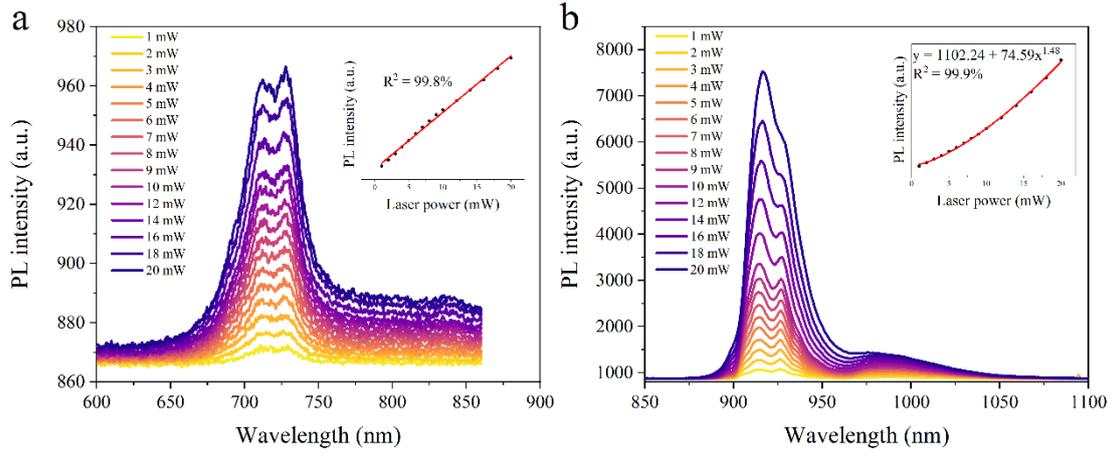

Figure 2 (a) Power-dependent PL spectra in the range of 600-850 nm, showing the high-energy $PL_1$ emission, inset shows a linear relationship between $PL_1$ peak intensity and excitation power with $R^2$ = 99.8%. (b) Power-dependent PL spectra in the range of 850-1100 nm, showing the infrared $PL_2$/$PL_3$ emissions, inset shows a non-linear relationship between $PL_2$ peak intensity and excitation power fitted with the equation $y = 1102.24 + 74.59x^{1.48}$ ($R^2$ = 99.9%).

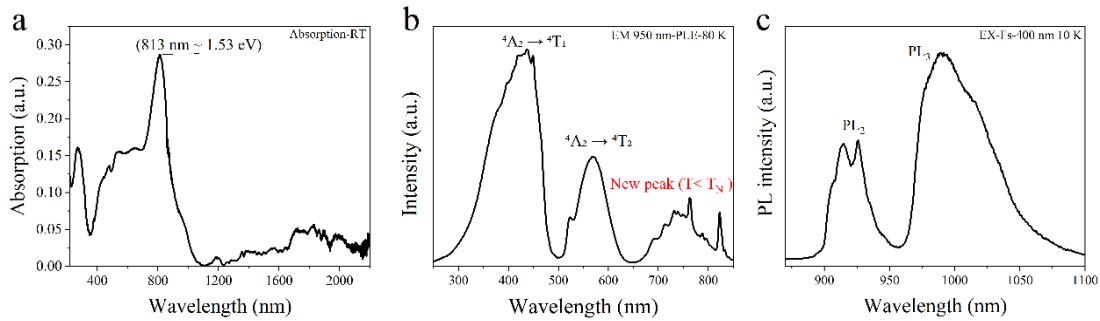

Figure 3 (a) Absorption spectrum of CrSBr measured at room temperature using a UV-Vis absorption spectrometer. (b) PLE spectrum monitored at 950 nm emission measured by HORIBA infrared detection system at 80 K (Note: two small peaks in the third peak region are instrumental artifacts). (c) PL spectrum excited by femtosecond laser (400 nm, 84.2 MHz, 5 mW) at 10 K, showing two distinct emission peaks $PL_2$ and $PL_3$ in the range of 890-1100 nm.

In temperature-dependent PL studies of CrSBr using 488 nm laser excitation (20 mW), we observed high-energy $PL_1$ emission persisting within 10-200 K, with its intensity reaching a maximum at 75 K, suggesting phonon involvement ($A_g^1$) in the $PL_1$ band transition. Above 75 K, the peak intensity gradually decreased, exhibiting a blue-shift at 100 K followed by a red-shift until stabilizing near 726 nm at 200 K (Fig S5a and Fig S5c). In the same excitation condition, $PL_2$ emission also was primarily influenced by phonons within 10-75 K(Fig S5b and Fig S5d), reaching maximum intensity at 75 K (corresponding to phonon energy of $A_g^1$) [31], which supply a direct evidence for its small excitonic polaron in single layer.

Notably, below 40 K, subtle variations in $PL_2/PL_3$ indicated the emergence of a new magnetic phase (Fig 4b-d), which could be attributed to either FM ordering in the bulk 2d crystal or a spin freezing process resulting from gradual reduction of spin fluctuations[25,32]. the former $PL_2$ could be attributed to the single layer FM magnetic polaronic phase out of $Cr^{3+}$ trimer, while the $PL_3$ can be assigned to the bipolaronic magnetic exciton formation in two neighboring layers via 2LO phobon[30,33,34]. This state indicates a new FM exciton in adjacent layer coupling, accompanied with a single $Cr^{3+}$ polaron nearby in single layer. The $PL_1$ exciton can coexist with $PL_3$ exciton, and reduction of $PL_1$ exciton can also destroy the $PL_3$ exciton. $PL_3$ occurrence is due to strong FM interaction, 2LO coupling and low phonon scattering. we employed 400 nm femtosecond laser excitation, which revealed $PL_2$ splitting into two components ($PL_2^1$ and $PL_2^2$) below 75 K (Fig 4b-c), with an energy separation of 120.56 cm$^{-1}$ at 17 K corresponding to the $A_g^1$ phonon mode, which is its polaronic character. particularly the intensity of both $PL_2^1$ and $PL_2^2$ decrease below 40 K (Fig 4c), because some of them divided into $PL_3$ and $PL_1$ in this temperature zone. As temperature increase, $PL_2$ intensity continued to rise to its maximum at 140 K (Fig 4b-d), coinciding with the FM transition temperature of monolayer CrSBr, strongly suggesting its trimer FM excitonic polaron (Fig 1a, three adjacent excitons along the *b-axis*) within a single layer. Significantly, $PL_3$ disappeared approaching 130 K (Fig 4a-b) with its corresponding $PLE_3$ vanishing at 130 K, demonstrating its characteristic as an FM polaronic exciton, because over 130 K more paramagnetic states start to occur in multilayers.

These findings were further corroborated by 800 nm femtosecond laser pulse excitation PL experiments, where $PL_2$ also gave its maximum emission at 140 K (Fig S8), but with much higher $PL_3$ dynamic emission than $PL_2$ at 15 K, reinforcing its association with monolayer and adjacent layer FM phase, while the disappearance of $PL_3$ above 130 K confirmed its characteristic as an interlayer FM exciton, caused by the initial AFM phase (Fig S8). In the bulk, interactions between $Cr^{3+}$ ions, bound with

$S^{2-}$ and $Br^-$ ions, create complicated electronic states, where the Cr-S-Cr and Cr-Br-Cr bond angles (~90° favoring FM coupling, ~160° leading to varied FM/AFM interactions) and covalency significantly influence their magnetic interactions in different direction and spacing; furthermore, our experiments revealed a 488.35 meV energy difference between PLE $^4T_2^{(1)}$ and $^4T_2^{(2)}$, while $PL_2$ and $PL_3$ showed 95.3 meV separation, suggesting direction and number dependent FM coupled aggregates for EMP rather than direct $^4T_2^{(1)}$ and $^4T_2^{(2)}$ separation. As layer number increases, phonon zone folding may contribute to exciton coherence and interlayer FM coupling can contribute to a more stable and longer-range magnetic polaron in an anisotropic structure and bonding, leading to varied EMPs and generating similar PLE bands, observable only below $T_N$; additionally, in the short and long AFM phase, the Raman zone folding effect may work in a different range. In this AFM case (10-132 K), Raman spectra showed dominant overtone modes at ($120 \pm 5$ cm$^{-1}$), ($252 \pm 5$ cm$^{-1}$), and ($546 \pm 5$ cm$^{-1}$), while below 132 K, second-order features at $461 \pm 5$ cm$^{-1}$, $546 \pm 5$ cm$^{-1}$, $646 \pm 5$ cm$^{-1}$, $746 \pm 5$ cm$^{-1}$ became increasingly pronounced, reflecting its very efficient local intralayer FM order with naturally enhanced electron-phonon coupling(Fig S7). Therefore, this is a long-range AFM, not blocking the local FM exciton to emit.

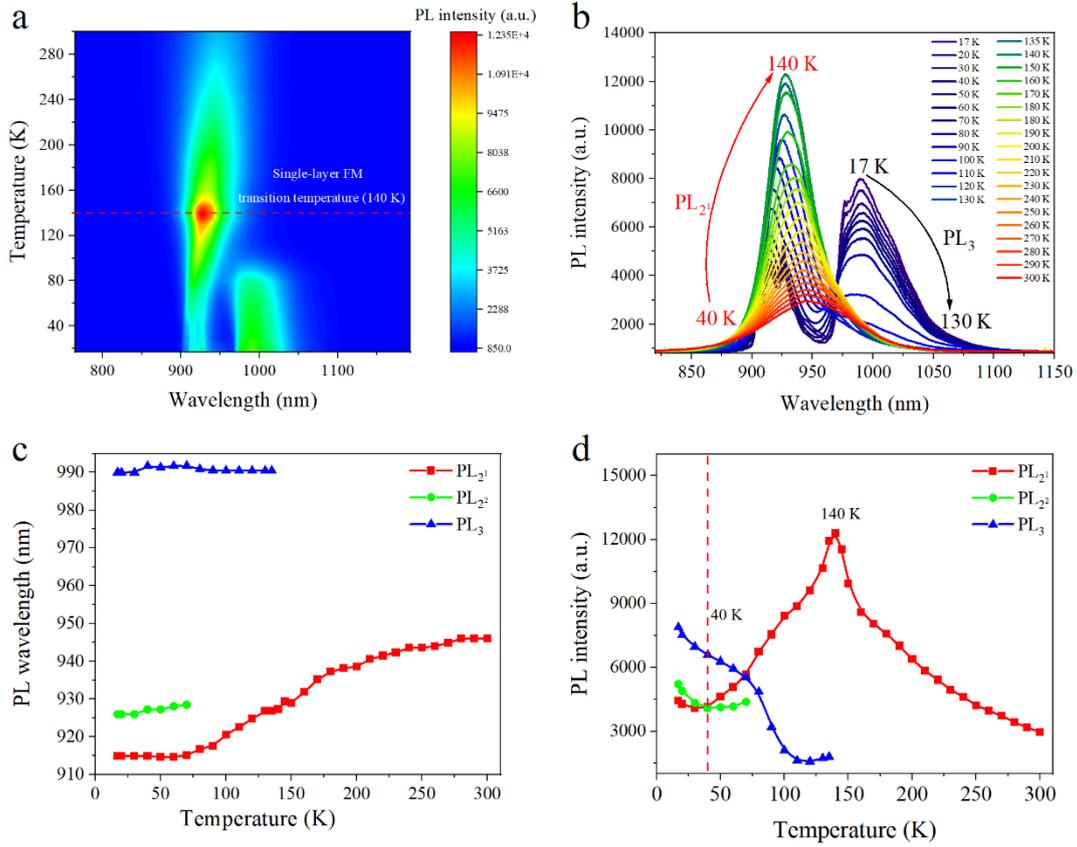

Figure 4 temperature-dependent PL measurements excited by femtosecond laser (400 nm, 84.2 MHz, 20 mW): (a) 2D pseudo-color map of $PL_2/PL_3$ from 17 K to 300 K, (b) $PL_2/PL_3$ spectra from 17 K to 300 K, (c) Temperature dependence of $PL_2/PL_3$ peak wavelengths, (d)Temperature dependence of $PL_2/PL_3$ peak intensities from 17 K to 300 K.

In bulk CrSBr, the complicated interactions between $Cr^{3+}$ ions, mediated by $S^{2-}$ and $Br^-$ ions, form a 2D vdW structure, lead to distinct polaronic emission. The high-energy $PL_1$ emission, previously overlooked due to its requirement for high-energy or ultraviolet excitation above band-gap[10,35,36], was observed here as an extremely weak emission near 720 nm at 488 nm laser excitation, persisting even at room temperature. PL measurements at room temperature using an FLS1000 spectrometer showed strongest $PL_1$ intensity at 315 nm excitation, consistent with the 325 nm laser excitation results (Fig. S1b). The millisecond-range lifetime of $PL_1$ (Fig S9b) and its temperature-dependent behavior (lifetime progressively increasing with temperature from 80-300 K due to phonon effect, Fig. S10c and S10d) strongly suggest its origin from $Cr^{3+}$ $^4T_2 \rightarrow {}^4A_2$ d-d transitions localized EMP due to coupling with LO phonon to relieve selection rule. In contrast, at PL$_2$ wavelength = 950 nm emission, showed lifetimes decreasing with increasing temperature at both nanosecond (Fig. S9d) and microsecond ranges, represented with the phonon scattering profile. This contrast indicates their phonon effect in different temperature range. The geometric configuration of Cr-S-Cr and Cr-Br-Cr bonds in layers significantly influences these magnetic interactions, with ~90° bond angles in layer favoring FM coupling. Our experiments revealed suggesting different FM coupled aggregates states. In this system, while the intralayer FM coupling persists, high-temperature interlayer magnetic coupling disappears and recover at low temperature, resulting in FM enhanced d-d transitions out of Cr-Cr pairs, trimers, tetramer or magnetic polarizers, whose absorption coefficients are large even below the band edge (~500 nm).

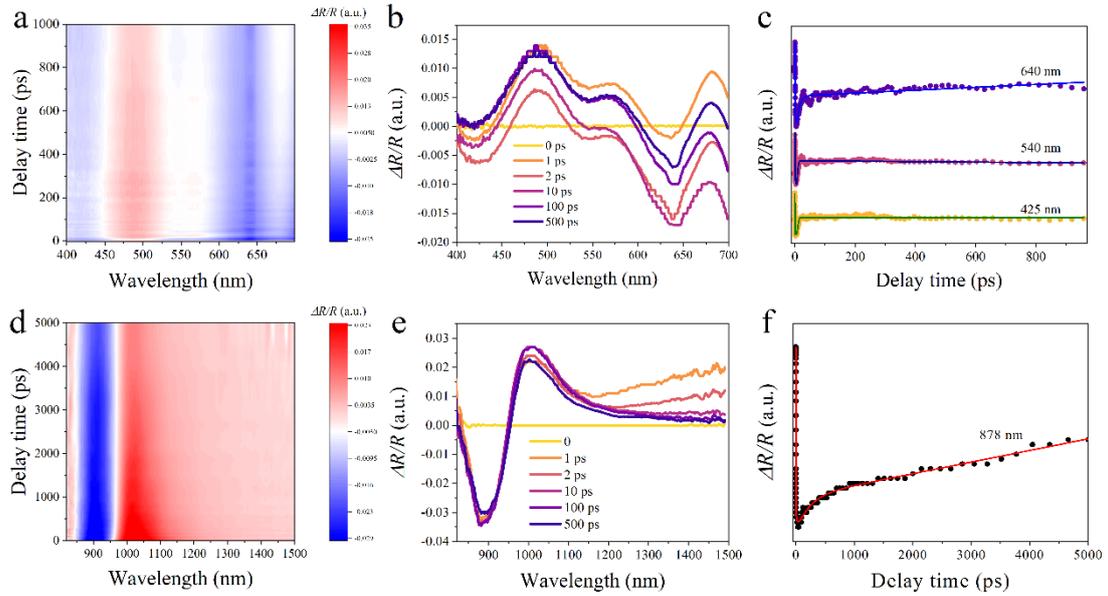

Figure 5 Transient absorption spectroscopy of bulk CrSBr excited by femtosecond laser (320 nm, 150 µW, beam spot ~100 µm) in reflection mode. (a) 2D pseudo-color plot of the temporal evolution of reflection spectrum in 400-700 nm range. (b) Transient reflection spectra at selected time delays in 400-700 nm range. (c) Extracted ΔR/R kinetics and fits at 425 nm, 540 nm, and 640 nm. (d) 2D pseudo-color plot of the temporal evolution of reflection spectrum in 875-1500 nm range. (e) Transient reflection spectra at selected time delays in 875-1500 nm range. (f) Extracted ΔR/R kinetics and fits at 878 nm.

Through femtosecond transient absorption (TA) spectroscopy studies, their excitonic dynamics and related microscopic interactions are unveiled in CrSBr systematically. In the visible region, three major ground-state positive absorption and bleaching signals were observed (Fig 5): the 425 nm positive signal corresponds to the biexcitons from the d-d transitions, exhibiting $\tau_1$ and $\tau_2$ fast processes that reflect complex excitation and thermalization of spin-coupled states with phonon interactions; which correlates with the PLE band at 472 nm and exhibits a clear electron-hole plasma relaxation profile in tens of nanoseconds scale, including contributions from Cr-S charge transfer band tail; the 500 nm bleaching imply its bandage character, while 540 nm signal corresponds to the $^4A_2 \rightarrow {}^4T_1$ transition ($\tau_1$=0.11 ps, $\tau_2$=0.64 ps, $\tau_3$=40.47 ns, Table S1). the 640 nm positive signal corresponds to the biexcitonic profile out of the $^4A_2 \rightarrow {}^4T_2$ transition ($\tau_1$=0.30 ps, $\tau_2$=0.68 ps, $\tau_3$=6.136 ns, Table S1), associated with the PLE band near 600 nm, with its lifetime characteristics indicating FM coupled exciton-magnetic polaron nature[37]. Notably, the high-energy excitons observed at 425 nm can contribute to $PL_1$ emission, which originates from the charge transfer between Cr-S or Br, and d-d transitions of coupled $Cr^{3+}$ ions. Its millisecond-scale long lifetime primarily stems from magnetic coupling between spin-down $Cr^{3+}$ ions and electron-phonon coupling. Below 500 nm, the crystal field effect can be significantly enhanced the transition oscillator strength via FM coupling, resulting in varied transient nonlinear optical absorption in this region. In the infrared range (Fig 5d-f), a pronounced bleaching band was observed at 878 nm (1.41 eV), that is corresponding to the lowest bandage out of the localized bipolaronic magnetic exciton out of the d-d transition of $Cr^{3+}$ pair or trimer and electron-phonon coupling. The fast dynamics observed in this near-infrared region (878-910 nm) ($\tau_1$=24.38 ps, $\tau_2$=226.79 ps, $\tau_3$=9.83 μs, Table S1) corresponds to dynamic exciton character. Mostly they originate from the aggregated states of three adjacent $Cr^{3+}$ ions in a FM layer and four $Cr^{3+}$ in inter-layer sites. Within the multi-layers, each two neighboring layers can convert their spins into opposite direction with AFM interaction through $S^{2-}$-mediated exchange., which produce the high temperature PL disappearance. The partially coherent exciton recombination observed at 950-1000 nm indicates that the FM exchange interaction among these aggregated $Cr^{3+}$ ions enhance electronic delocalization and propagation via phonon coupling effect. allowing small polaron exciton and large bipolaronic exciton to form dynamically and recombine to emit light in the FM zones, that account for $PL_2$ with a shorter microsecond-scale lifetime as compared to $PL_1$.

**Magneto-PL**

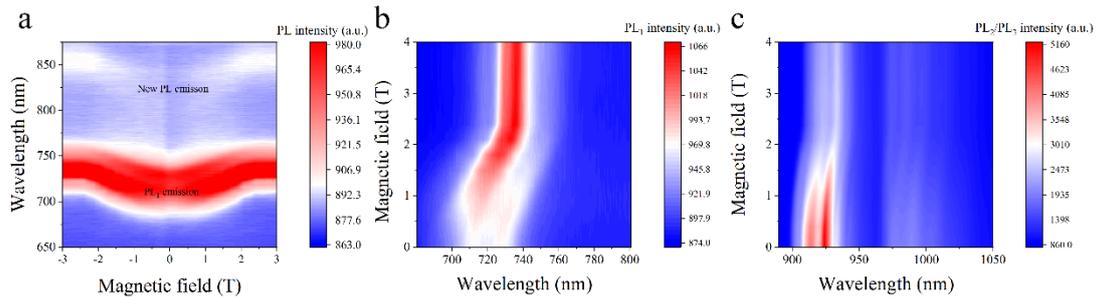

Figure 6 (a) Magneto-PL measurements at 10 K under 488 nm laser excitation (20 mW) with magnetic field applied perpendicular to the a-b plane, showing $PL_1$ and emergence of a new PL peak at 850 nm (dark state exciton) under strong magnetic field. (b) Magnetic field-dependent PL spectra (0-4 T) in the high-energy region showing $PL_2$ evolution. (c) Magnetic field-dependent PL spectra (0-4 T) in the infrared region showing $PL_2/PL_3$ evolution.

Fig.6 presented the magnetic field-dependent PL results obtained in bulk CrSBr under 488 nm excitation at 10 K. At 0.4 T, both $PL_1$ and $PL_2/PL_3$ emissions showed red-shifts, but $PL_1$ intensity increased while $PL_2/PL_3$ decreased, attributed to their different magnetic polaron dipole orientations. $PL_1$ dipole get stronger when the magnetic field perpendicular to the layer rise, while the $PL_2/PL_3$ emission band reduce. $PL_2/PL_3$ behavior at different magnetic fields under 400 nm fs pulse excitation remains consistent with 488 nm CW laser excitation, confirming the intrinsic nature of the magnetic polaron response (Fig S14a). Critical magnetic field transitions occurred at 1 T and 2.1 T, where the many-body effects lead to competition among magnetic field related exciton dipole anisotropy, exchange energy, and dipole energy, reflected in the clear spectral changes[38]. From 0.4 T to 1 T, $PL_1$ and the newly observed $PL_{1a}$ (850 nm corresponding dark state magnetic biexciton) bands showed gradual peak wavelength increases, followed by rapid surge and stabilization at 2.1 T (Fig 6a), representing their complete spin and polaron dipole transformation to the magnetic field. This magnetic field-induced PL intensity variation reveals the population competition between different excitonic states: under stronger magnetic fields in perpendicular to layer, $PL_2/PL_3$ excitons partially dissociate into single $Cr^{3+}$ ions and dark state pair excitons for higher emission. The emission band FWHM competitive nature is also in Fig S12b-c[38]. At room temperature under magnetic fields, three magneto-polarized peaks appeared on the long wavelength band side in Fig S13, with energy spans of 223.75 cm$^{-1}$, 299.74 cm$^{-1}$, and 523.49 cm$^{-1}$, indicative of the spin-2LO phonon-exciton coupling character in the enhanced emission processes just out of those bipolaronic magnetic exciton in this compound. The observed PL peak separation of 223.75 cm$^{-1}$ agrees well with the calculated double phonon-magnon effect ($2\times A_g^1$ + magnon ~224.07 cm$^{-1}$), previous literature reported magnon frequencies of 24 GHz and 34 GHz (0.8 cm$^{-1}$ and 1.13 cm$^{-1}$)[13,14], further supporting the existence of EMP at room temperature.

**DFT calculations**

To verify the microscopic origin of multiple exciton species observed in our experiments, we use the density functional theory (DFT) calculations with the generalized gradient approximation of the Perdew–Burke–Ernzerhof (PBE)[39,40] hybrid functional and DFT+U correction (U = 4 eV) revealed three distinct $Cr^{3+}$ configurations contributing to the different optical transitions and the vdW Material Amendment D3G[41,42]: single $Cr^{3+}$ ions (6-atom cell) showed a band gap of 2.28 eV (Fig S15) with random spins, where the conduction band is dominated by $Cr^{3+}$ d-orbitals and valence band by $Br^-$ and $S^{2-}$ orbitals, getting $PL_1$ emission from spin-down single $Cr^{3+}$ ions. $Cr^{3+}$ trimers (24-atom supercell) are obtained at 1.54 eV (Fig S16a, *b-axis*) and 1.57 eV (Fig S16b *a-axis*) for FM spin-up alignment, with a small fraction showing 2.0 eV (*b-axis*) and 2.2 eV (*a-axis*) values for spin-down states, corresponding to $PL_1$ emission from single $Cr^{3+}$ ion within monolayer; and the FM $Cr^{3+}$ trimer is consistent with $PL_2$ emission within single layers. The non-collinear calculations for $Cr^{3+}$ tetramers (12-atom bilayer supercells) revealed values of 1.452 eV (Fig S17a *b-axis*) and 1.458 eV (Fig S17 b *a-axis*), supporting a $PL_3$ band that origin from interlayer excitons only in multilayer structures. This result indicate that in the AFM phase there are two FM layers adjacent to another two layers in the whole structures[17]. The incorporation of spin-orbit coupling significantly affected its band structure near high symmetry points, consistent with observed optical responses in magnetic field, validating that above PL bands mainly originates from single $Cr^{3+}$ ion, trimers, tetramer all in *b-axis* in this structure, thereby providing a comprehensive understanding on the multi-body effect in the magnetic and optical properties of 2D CrSBr structure.

## Conclusion

In summary, we have systematically investigated the complicated interaction among the spin-phonon-exciton in CrSBr through comprehensive spectroscopic studies and theoretical calculations. We identify three distinct radiation relaxation channels out of local magnetic polarons within bandgap: $PL_1$ (~720 nm) originating from the d-d transition of a single $Cr^{3+}$ ion with millisecond lifetimes in the multilayer compound; $PL_2$ (~920 nm) originating from an FM coupled trimer within a monolayer due to the exciton-phonon interaction with FM coupled spins, i.e., a local excitonic magnetic polaron; and $PL_3$ (~990 nm) originating from an interlayer FM exciton in a two-layer structure, due to bipolaronic magnetic coupling via two-phonon. A new dark-state magnetic excitons ($PL_{1a}$) at about 850nm appear in the perpendicular magnetic field to layer. their PL intensities change with their dipole stability and dipole orientation in magnetic field and low temperature. $PL_2$ reaches a maximum in the monolayer ferromagnetic transition temperature (140 K), while $PL_3$ disappears above the bulk $T_N$ temperature (132 K). The band structure and local magnetic polaronic band can be identified by the TA absorption profile. DFT calculations confirm the above experimental findings on the different $Cr^{3+}$ aggregation and optical transitions in this

2D structure. the excellent 2d magnetic polaronic exciton stability and efficient magnetic field responses in CrSBr make it a promising platform for future quantum information technologies based on the spin-phonon-exciton interactions.

# Methods

## Crystal growth

CrSBr bulk single crystals were synthesized using a chemical vapor transport method. Chromium (99.99%, -60 mesh;), sulfur (99.9999%, Aladdin Reagent Network), and bromine (99.9999%; Nanning Blue Sky Experimental Equipment Co, Ltd.) were placed in a quartz glass ampoule with a general stoichiometry of approximately 1:1:1.1. Chromium powder, sulfur, and liquid bromine were added into a quartz glass tube. The lower end of the glass tube was vacuum sealed by placing it in a liquid nitrogen bath. Initially, the sample was preheated for 10 hours in a dual-temperature zone furnace with the source zone at 700 °C and the low-temperature zone at 360 °C. Subsequently, the tube was inspected for any residual liquid bromine. Once no liquid bromine was detected, the firing process commenced with the source zone at 700 °C and the growth zone at 900 °C, lasting for 25 hours. Within 1 hour, the temperatures were reversed with the source zone at 900 °C and the growth zone at 850 °C. Over the next 7 days, the temperature in the source zone gradually increased from 900 °C to 940 °C, while the temperature in the growth zone decreased from 850 °C to 800 °C. Finally, the system was allowed to cool naturally. Our sample was successfully synthesized after several days of employing chemical vapor transport. Acetone and anhydrous alcohol were used to clean any residues on the surface of the raw materials, thereby obtaining the desired bulk samples. In this experiment, bulk samples were primarily used, with dimensions of approximately 4×1.5×0.2 mm³

**Optical spectroscopy**

The photoluminescence spectroscopy testing system is from WITec's Raman testing system. The excitation wavelength uses a continuous, variable average power laser with a wavelength of 488 nm for excitation. The sample is placed in a low-temperature system using a closed-cycle cryostat from Sumitomo, USA. The cryostat has a minimum temperature of 10 K and is equipped with a 7 T solenoid magnet. The solenoid magnet is placed inside a separate closed-cycle cooling system. The direction of the magnetic field is perpendicular to the surface of the sample. The scanning electron microscopy (SEM, Hitachi SU8020) was used to observe the morphology. The energy-dispersive spectrometry (EDS, Oxford X-Max Aztec) was used to collect the element composition and distribution. And the steady-state S3 PL spectra, temperature-dependent PLE spectra and temperature-dependent time-resolved PL (TRPL) were measured with a Horiba Jobin Yvon Fluorolog$^{-3}$ spectrometer. TRPL was also collected using Edinburgh FLS1000 fluorescence spectrometer. The electronic absorption spectra were measured using a UV-VIS-NIR spectrophotometer (PerkinElmer Instruments, Lambda 750). The femtosecond laser transient absorption system used in this article comes from Dalian Chuangrui Spectroscopy Company in China. The femtosecond laser and amplification system are from the American Spectral Physics Company. The pulse width is 100 fs and the wavelength is 800 nm. After the amplification stage output the average power is above 4 W and the repetition frequency is 1000 Hz. After being amplified and output by the amplification stage, a beam of laser passes through the TOPAS optical parametric amplifier (OPA) amplification system as pump laser. A beam of laser is focused on sapphire ($Al_2O_3$) or calcium fluoride ($CaF_2$) through the reflection system to produce chirped white light. The Idler signal light is generated by TOPAS and then amplified to obtain from 200 - 2000 nm pump laser. In this article, we use 320 nm pump light with an average power of 150 μw. The diameter spot size focused on the sample is about 100 μm. The probe light is perpendicular to the sample plane. The angle between the probe light and the pump laser is about 11°. The reflection system introduces the signal light focusing fiber into the detector and detects changes in reflectivity through the delay system.

## Acknowledgments

The authors thank the Scientific and Technological Bases and Talents of Guangxi (Guike AD23026119, AD21238027), the Guangxi NSF project (2020GXNSFDA-238004) and the "Guangxi Bagui Scholars" foundation for financial support.

## Reference

1   Ghiasi, T. S. *et al.* Nitrogen-vacancy magnetometry of CrSBr by diamond membrane transfer. *npj 2D Materials and Applications* **7**, doi:10.1038/s41699-023-00423-y (2023).
2   Ghiasi, T. S. *et al.* Electrical and thermal generation of spin currents by magnetic bilayer graphene. *Nat Nanotechnol* **16**, 788-794, doi:10.1038/s41565-021-00887-3 (2021).
3   Jo, J. *et al.* Local control of superconductivity in a NbSe(2)/CrSBr van der Waals heterostructure. *Nat Commun* **14**, 7253, doi:10.1038/s41467-023-43111-7 (2023).
4   Ruta, F. L. *et al.* Hyperbolic exciton polaritons in a van der Waals magnet. *Nature Communications* **14**, 8261 (2023).
5   Wang, C. *et al.* A family of high-temperature ferromagnetic monolayers with locked spin-dichroism-mobility anisotropy: MnNX and CrCX (X= Cl, Br, I; C= S, Se, Te). *Science Bulletin* **64**, 293-300 (2019).
6   Wilson, N. P. *et al.* Interlayer electronic coupling on demand in a 2D magnetic semiconductor. *Nature Materials* **20**, 1657-1662 (2021).
7   Du, L. New excitons in multilayer 2D materials. *Nature Reviews Physics*, doi:10.1038/s42254-024-00704-5 (2024).
8   Telford, E. J. *et al.* Coupling between magnetic order and charge transport in a two-dimensional magnetic semiconductor. *Nat Mater* **21**, 754-760, doi:10.1038/s41563-022-01245-x (2022).
9   Bae, Y. J. *et al.* Exciton-coupled coherent magnons in a 2D semiconductor. *Nature* **609**, 282-286, doi:10.1038/s41586-022-05024-1 (2022).

10 Cham, T. M. J. *et al.* Exchange Bias Between van der Waals Materials: Tilted Magnetic States and Field-Free Spin–Orbit-Torque Switching. *Advanced Materials* **36**, 2305739 (2024).

11 López-Paz, S. A. *et al.* Dynamic magnetic crossover at the origin of the hidden-order in van der Waals antiferromagnet CrSBr. *Nature Communications* **13**, 4745 (2022).

12 Telford, E. J. *et al.* Layered Antiferromagnetism Induces Large Negative Magnetoresistance in the van der Waals Semiconductor CrSBr. *Adv Mater* **32**, e2003240, doi:10.1002/adma.202003240 (2020).

13 Diederich, G. M. *et al.* Tunable interaction between excitons and hybridized magnons in a layered semiconductor. *Nat Nanotechnol* **18**, 23-28, doi:10.1038/s41565-022-01259-1 (2023).

14 Dirnberger, F. *et al.* Magneto-optics in a van der Waals magnet tuned by self-hybridized polaritons. *Nature* **620**, 533-537, doi:10.1038/s41586-023-06275-2 (2023).

15 Ruta, F. L. *et al.* Hyperbolic exciton polaritons in a van der Waals magnet. *Nat Commun* **14**, 8261, doi:10.1038/s41467-023-44100-6 (2023).

16 Wilson, N. P. *et al.* Interlayer electronic coupling on demand in a 2D magnetic semiconductor. *Nat Mater* **20**, 1657-1662, doi:10.1038/s41563-021-01070-8 (2021).

17 Sun, Z. *et al.* Resolving and routing the magnetic polymorphs in 2D layered antiferromagnet. *arXiv preprint arXiv:2410.02993* (2024).

18 Lin, K. *et al.* Probing the Band Splitting near the Γ Point in the van der Waals Magnetic Semiconductor CrSBr. *The Journal of Physical Chemistry Letters* **15**, 6010-6016 (2024).

19 Jia, Z. *et al.* Strategies to approach high performance in Cr(3+)-doped phosphors for high-power NIR-LED light sources. *Light Sci Appl* **9**, 86, doi:10.1038/s41377-020-0326-8 (2020).

20 Liu, G. *et al.* Laser-driven broadband near-infrared light source with watt-level output. *Nature Photonics* **18**, 562-568, doi:10.1038/s41566-024-01400-7 (2024).

21 Liu, S., Du, J., Song, Z., Ma, C. & Liu, Q. Intervalence charge transfer of Cr(3+)-Cr(3+) aggregation for NIR-II luminescence. *Light Sci Appl* **12**, 181, doi:10.1038/s41377-023-01219-x (2023).

22 Lin, K. *et al.* Strong Exciton-Phonon Coupling as a Fingerprint of Magnetic Ordering in van der Waals Layered CrSBr. *ACS Nano* **18**, 2898-2905, doi:10.1021/acsnano.3c07236 (2024).

23 Marques-Moros, F., Boix-Constant, C., Manas-Valero, S., Canet-Ferrer, J. & Coronado, E. Interplay between Optical Emission and Magnetism in the van der Waals Magnetic Semiconductor CrSBr in the Two-Dimensional Limit. *ACS Nano* **17**, 13224-13231, doi:10.1021/acsnano.3c00375 (2023).

24 Cenker, J. *et al.* Reversible strain-induced magnetic phase transition in a van der Waals magnet. *Nat Nanotechnol* **17**, 256-261, doi:10.1038/s41565-021-01052-6 (2022).

25 Klein, J. *et al.* Sensing the Local Magnetic Environment through Optically Active Defects in a Layered Magnetic Semiconductor. *ACS Nano* **17**, 288-299, doi:10.1021/acsnano.2c07655 (2023).

26 Wang, T. *et al.* Magnetically-dressed CrSBr exciton-polaritons in ultrastrong coupling regime. *Nat Commun* **14**, 5966, doi:10.1038/s41467-023-41688-7 (2023).

27 Liu, R., Shi, L. & Zou, B. Magnetic exciton relaxation and spin–spin interaction by the time-

28  Pankove, J. I. *Optical processes in semiconductors*.   (Courier Corporation, 1975).

29  Li, C. *et al.* 2D CrSBr Enables Magnetically Controllable Exciton-Polaritons in an Open Cavity. *Advanced Functional Materials*, doi:10.1002/adfm.202411589 (2024).

30  Huang, T. *et al.* Magnetic polaronic and bipolaronic excitons in Mn(II) doped (TDMP)PbBr4 and their high emission. *Nano Energy* **93**, 106863, doi:10.1016/j.nanoen.2021.106863 (2022).

31  Ge, S. *et al.* Photon and Phonon Coherence to Enhance Photoluminescence by Magnetic Polarons in Mn-Doped Rb3Cd2Cl7 Perovskites. *The Journal of Physical Chemistry C* **126**, 18855-18866, doi:10.1021/acs.jpcc.2c05179 (2022).

32  Pawbake, A. *et al.* Raman scattering signatures of strong spin-phonon coupling in the bulk magnetic van der Waals material CrSBr. *Physical Review B* **107**, doi:10.1103/PhysRevB.107.075421 (2023).

33  Devreese, J. T. & Alexandrov, A. S. Fröhlich polaron and bipolaron: recent developments. *Reports on Progress in Physics* **72**, 066501 (2009).

34  Emin, D. & Hillery, M. S. Continuum studies of magnetic polarons and bipolarons in antiferromagnets. *Physical Review B* **37**, 4060 (1988).

35  Lee, K. *et al.* Magnetic order and symmetry in the 2D semiconductor CrSBr. *Nano Letters* **21**, 3511-3517 (2021).

36  Lin, K. *et al.* Strong Exciton–Phonon Coupling as a Fingerprint of Magnetic Ordering in van der Waals Layered CrSBr. *ACS nano* **18**, 2898-2905 (2024).

37  Zou, S. *et al.* Bosonic lasing from collective exciton magnetic polarons in diluted magnetic nanowires and nanobelts. *ACS Photonics* **3**, 1809-1817 (2016).

38  Hänggi, P., Talkner, P. & Borkovec, M. Reaction-rate theory: fifty years after Kramers. *Reviews of modern physics* **62**, 251 (1990).

39  Kresse, G. & Furthmüller, J. Efficient iterative schemes for ab initio total-energy calculations using a plane-wave basis set. *Physical review B* **54**, 11169 (1996).

40  Perdew, J. P., Burke, K. & Ernzerhof, M. Generalized gradient approximation made simple. *Physical review letters* **77**, 3865 (1996).

41  Datta, B. *et al.* Magnon-mediated exciton-exciton interaction in a van der Waals antiferromagnet. *arXiv preprint arXiv:2409.18501* (2024).

42  Shi, J. *et al.* Giant Magneto-Exciton Coupling in 2D van der Waals CrSBr. *arXiv preprint arXiv:2409.18437* (2024).

# Supporting information

We have successfully synthesized high-quality CrSBr samples using the Chemical Vapor Transport (CVT) method (Fig S1d). This method involves the transport of solid raw materials from a high-temperature region to a lower-temperature region within a sealed environment, facilitating the growth of the material. A portion of the sample was subjected to a series of tests. The sample exhibits an average thickness of 68.41μm, a width of approximately 1.5 mm, and a length nearing 8 mm. These parameters were obtained by measuring the middle area of the sample using a profilometer (Fig S1c). To gain a deeper understanding of the sample's composition, we employed Energy-Dispersive Spectroscopy (EDS), a technique used under an electron microscope to analyze the chemical composition of materials. Through EDS, we discovered that the atomic ratios of Cr, S, and Br in the sample are 33.51%:32.66%:33.83%, respectively as shown in Figure S2, which closely aligns with the ideal ratio of 1:1:1. This suggests that our sample exhibits excellent chemical uniformity. In order to further explore the formation of excitons, we used femtosecond laser transient absorption spectroscopy for reflection testing. As shown in the figure, multiple absorption bands appeared at 400-700 nm. The data were extracted and fitted to obtain Table 1. Fitting formula as follows:

$$\Delta R(t)/R_0 = R_1 + R_2 + R_3 + y_0 = A_1 e^{(-t/\tau_1)} + A_2 e^{(-t/\tau_2)} + A_3 e^{(-t/\tau_3)} + y_0$$

This is a tri-exponential fitting function for transient absorption/reflectance spectral analysis. It contains three exponential terms, each representing a different lifetime component, where $A_1$, $A_2$, and $A_3$ are the amplitudes of the components, and $\tau_1$, $\tau_2$, and $\tau_3$ are the corresponding lifetimes. $y_0$ term denotes the baseline correction, i.e., the signal offset or final level. This model is capable of describing multiple relaxation processes occurring in the sample after photoexcitation, with each exponential term corresponding to a specific kinetic process.

**Details of theoretical calculation.**

All calculations at density functional theory are carried out using ABACUS

(Atomic-orbital Based Ab-initio Computation at UStc) with the projector augmented wave (PAW) method[1]. The generalized gradient approximation of Perdew-Burke-Ernzerhof (PBE) [2,3] functional with projector augmented wave method was employed. A plane-wave cutoff energy of 500 eV. The DFT+U method with U = 4 eV was applied to the Cr 3d orbitals to account for the strong electron correlation[4,5]. Van der Waals interactions were included through the D3 method with Grimme corrections(D3G). The energy convergence criterion was set to $1\times10^{-7}$ eV per atom. Band structures were calculated along the high-symmetry path Γ-X-S-Y-Γ. The density of states was calculated with an energy resolution of 0.05 eV in the range of -10 to 10 eV. The calculations cover a single $Cr^{3+}$ ion (6-atom cell) and three $Cr^{3+}$ ion (24-atom supercell) configurations aligned along the b-axis and a-axis using $5 \times 5 \times 5$ and $17 \times 5 \times 1$ k-point grids, with plane-wave truncation energies of 500 eV. The results show that the bandgap in the single cell when the spins are opposite along the b-axis is 2.28 eV, with the conduction band mainly contributed by the d orbitals of $Cr^{3+}$ and the valence band from the $Br^-$ and $S^{2-}$ orbitals[5-7]. In the trimer configuration, the band gaps are 1.54 eV and 1.57 eV when the spins are up along the b- and a-axis, respectively, and a small amount of the band gaps are about 2.0 eV (b-axis) and 2.2 eV (a-axis) when the spins are opposite[5]. In the trimer configuration, the conduction and valence bands are mainly contributed by the d-orbitals of $Cr^{3+}$ at spin-up, whereas the valence bands are mainly derived from the $Br^-$ and $S^{2-}$ orbitals at spin-down. The introduction of spin-orbit coupling (SOC) results in significant changes in the energy band structure, especially the splitting of the bands near the high symmetry point, slightly reduces the band gap value and affects the density of states details, especially near the Fermi energy level. This SOC effect is crucial for understanding the fine electronic structure of the material and possible optical leaps. Based on our calculations, we hypothesize that the PL at room temperature originates mainly from the trimer configuration in the b-axis direction, whereas the PL observed when a small amount of spin is reversed (spin down) may come from a single $Cr^{3+}$ ion. This speculation is consistent with the experimental observation of both $PL_1$ (spin down) and $PL_2$ (spin up) using 488 nm excitation light. These findings not only reveal the complex electronic structure and magnetic behavior in CrSBr, but also indicate that its optical properties strongly depend on the spin configuration and arrangement of the $Cr^{3+}$ ion. In our experiments, we could not observe luminescence in $PL_3$ with fewer layers, and hypothesized that this might be related to the aggregation state of the layers, assuming that interlayer ferromagnetism might occur with an increase in the number of layers, by arranging the two $Cr^{3+}$ between the upper and lower layers towards the a-axis or the b-axis, and the bilayer tetramer. Since it is an upper and lower layer, non-collinear calculations were used for the tetramer (12-atom supercells) as well as the 17 × 7 × 7 and 9 × 13 × 7 k-point lattices along the b- and a-axes, with other parameters unchanged. The results show that: (a) the spin noncollinearity of the tetramer bandgap toward the b-axis is 1.452 eV; (b) the spin noncollinearity of the tetramer bandgap toward the a-axis is 1.458 eV;

Table S1 uses the formula to fit a pair of data to obtain the lifetimes of the four ground-state bleaching signals:

| Wavelength | $\tau_1$ | $\tau_2$ | $\tau_3$ | $R^2$ |
| --- | --- | --- | --- | --- |
| 425 nm | 1.8584 ps | 7.3438 ps | 9E7 ps | 0.998 |
| 540 nm | 0.10648 ps | 0.64376 ps | 40473.049 ps | 0.998 |
| 640 nm | 0.29531 ps | 0.68098 ps | 6136.480 ps | 0.998 |
| 878 nm | 24.38085 ps | 226.7912 ps | 9828961.93 ps | 0.998 |

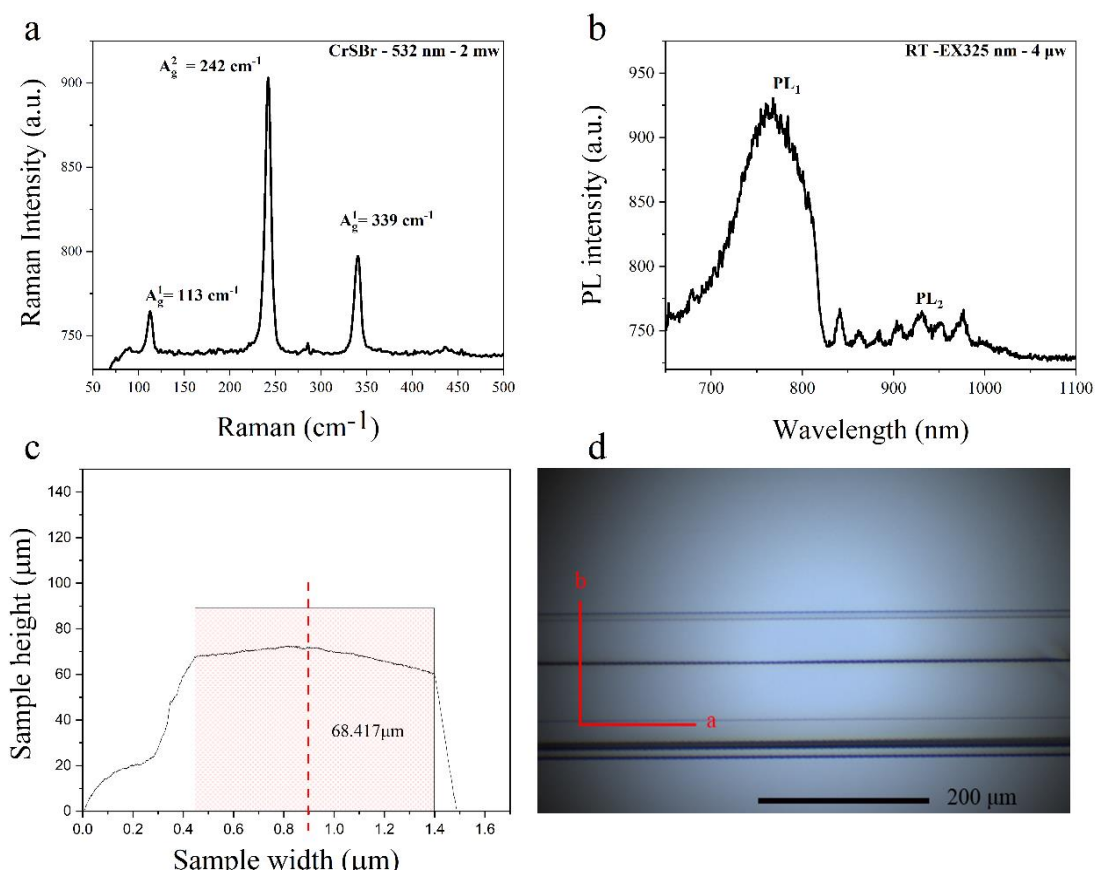

Figure S1 (a) the average thickness of the middle part of the sample is about 68.41 μm and the corresponding width value is 1.09 mm. (b) Microscopic image observed under a 10 X microscope in a low-temperature strong magnetic system.

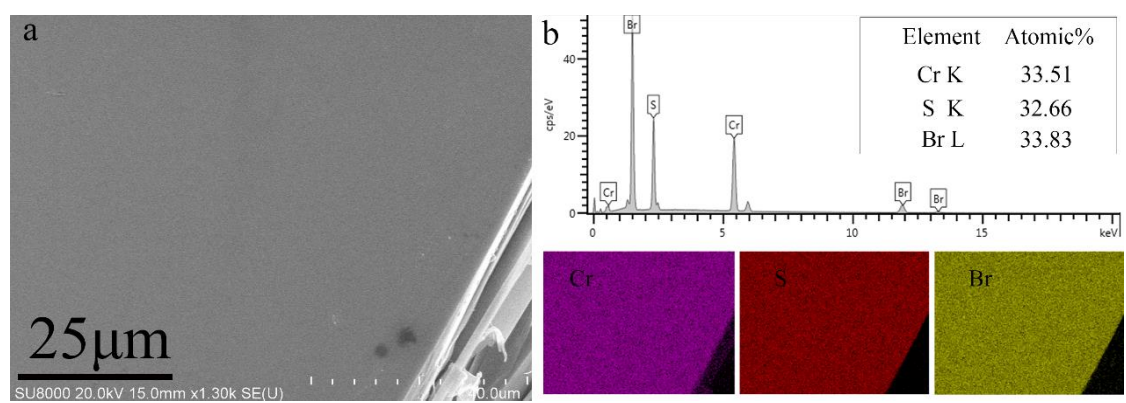

Figure S2 (a) The morphology of the sample was observed using a thermal field emission scanning electron microscope. (b) The proportion of elements in EDS and the element distribution map in the corresponding area.

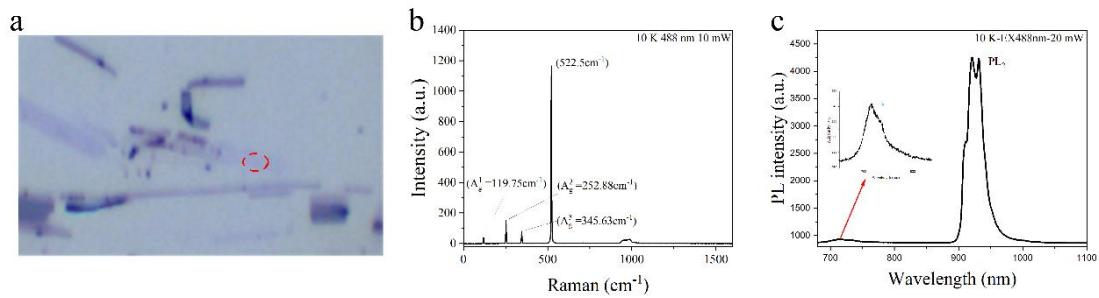

Figure. S3 (a) Microscopic magnification of near-monolayer CrSBr, highlighted within the red circle, as observed under a laser confocal microscope after its transfer onto the silicon wafer using the polydimethylsiloxane (PDMS) stamp technique. (b) is the corresponding Raman peaks in the red circle at 10 K. $A_g^1$= 119.75 cm$^{-1}$, $A_g^2$= 252.88 cm$^{-1}$, $A_g^3$= 345.63 cm$^{-1}$, which are very weak, with a very strong Raman peak of silicon is located at 522.5 cm$^{-1}$ intensity is 7.8-fold enhancement that of $A_g^2$, which further proves the state of close to one layer, (c) red region intrinsically corresponds to the spectrograms of PL$_1$ and PL$_2$, while PL$_3$ is almost disappeared.

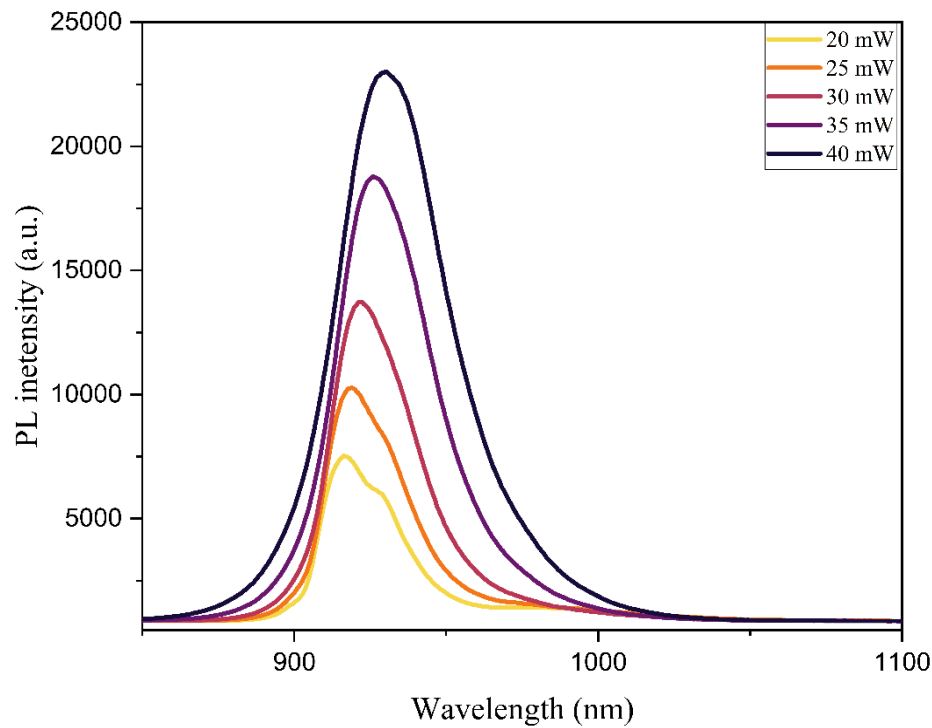

Figure S4 PL$_2$/PL$_3$ spectra at 10 K, 0T, 488 nm laser power from 20-40mW

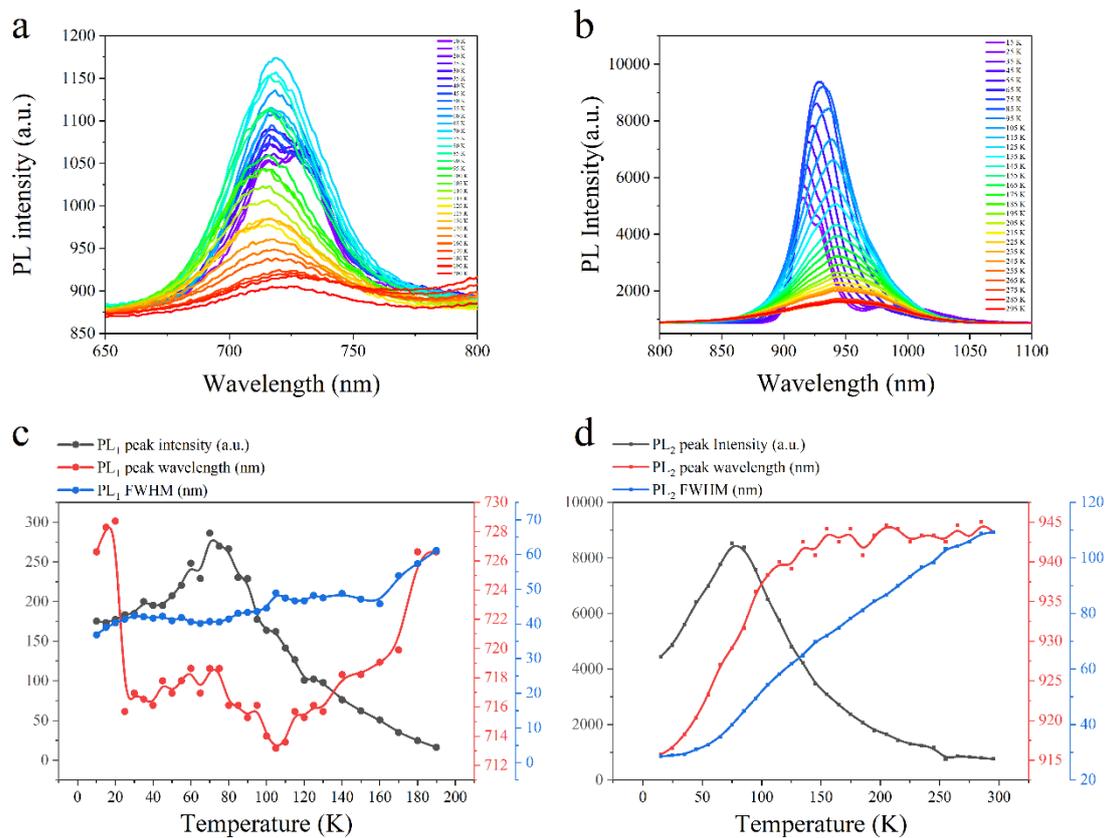

Figure S5 under 488 nm linearly polarized laser excitation with an average laser power of 20 mW, (a) temperature-dependent $PL_1$ spectrum in the range of 10-200 K, (b) temperature-dependent $PL_2/PL_3$ spectrum in the range of 10-300 K, (c) is a curve chart of the high-energy $PL_1$ peak at around 720 nm. As the temperature changes, (c) is a curve chart of the infrared $PL_2$ peak near 920 nm as the temperature changes.

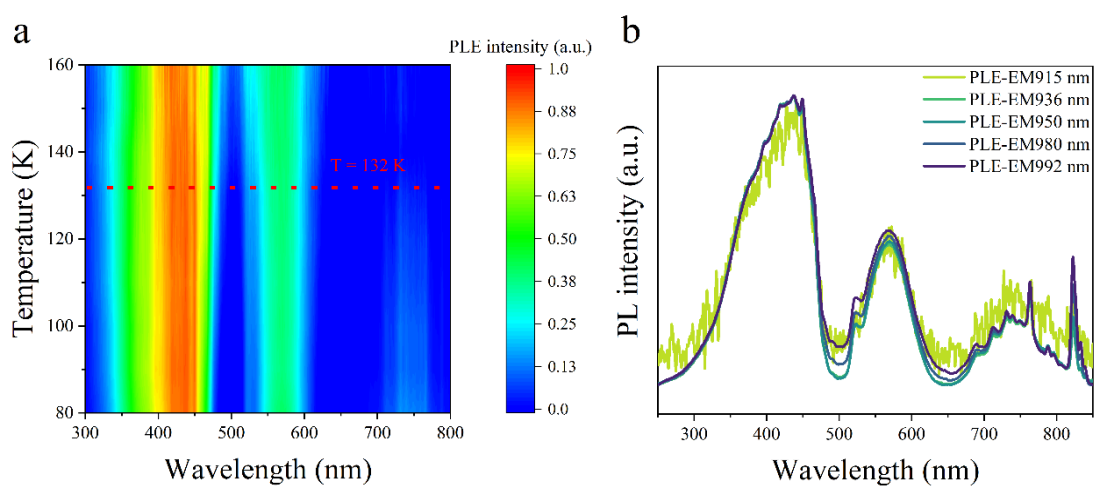

Figure S6 (a) PLE spectrum normalized from 80-160 K detected at 950 nm using a HORIBA infrared detector, (b) PLE spectrum normalized at 77 K with emission at 915 nm, 936 nm, 950 nm, 980 nm and 992 nm (the two small peaks on the third peak come from the measurement facility).

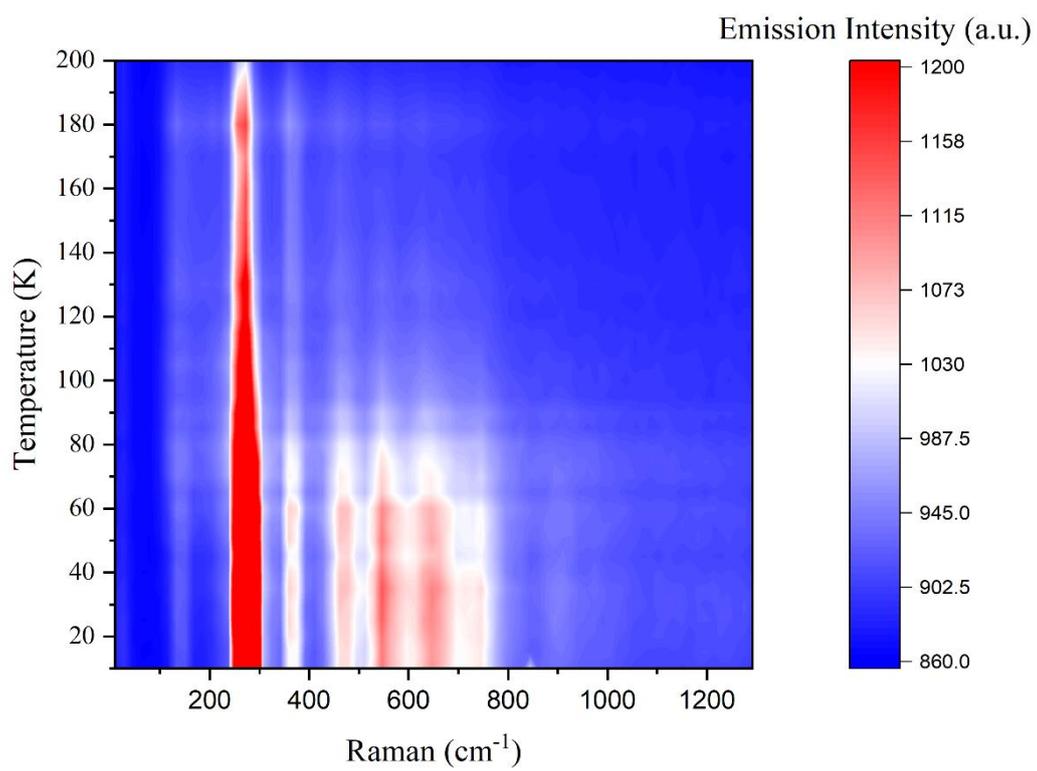

Figure S7 2D pseudo-color image of the Raman spectrum variation within 10 -1250 cm$^{-1}$ at 10-200 K.

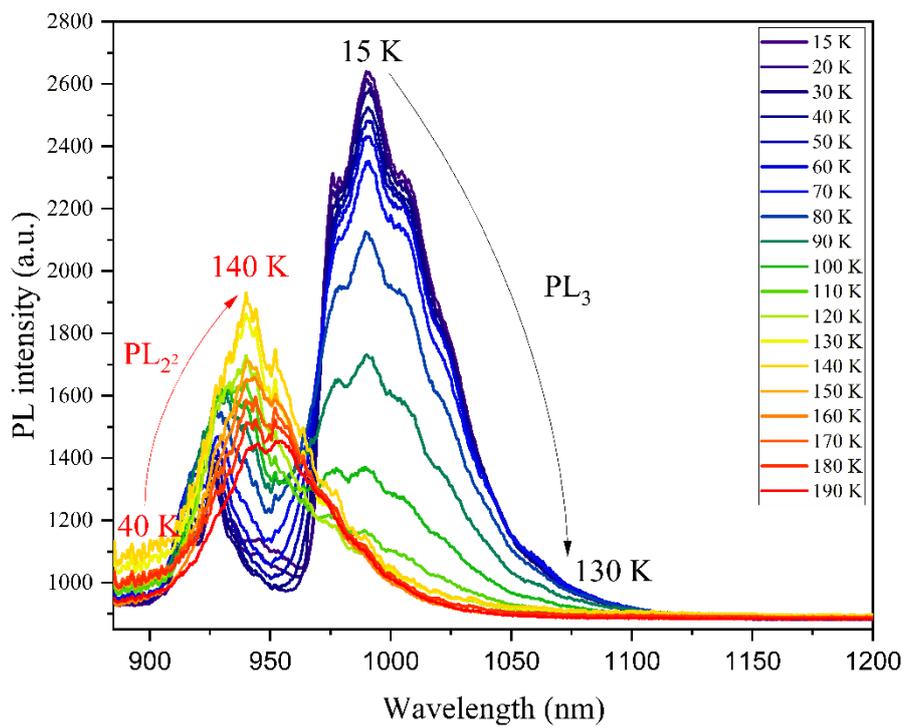

Figure S8 the temperature-dependent spectra at 800 nm, 84.2 MHZ, 100 fs, and 5.6 mW show that the peak intensity of $PL_2$ gradually increases at 40-140 K, with a maximum at 140 K, and that of $PL_3$ gradually decreases with the increase in temperature from 15 to 130 K. $PL_3$ disappears near 130 K.

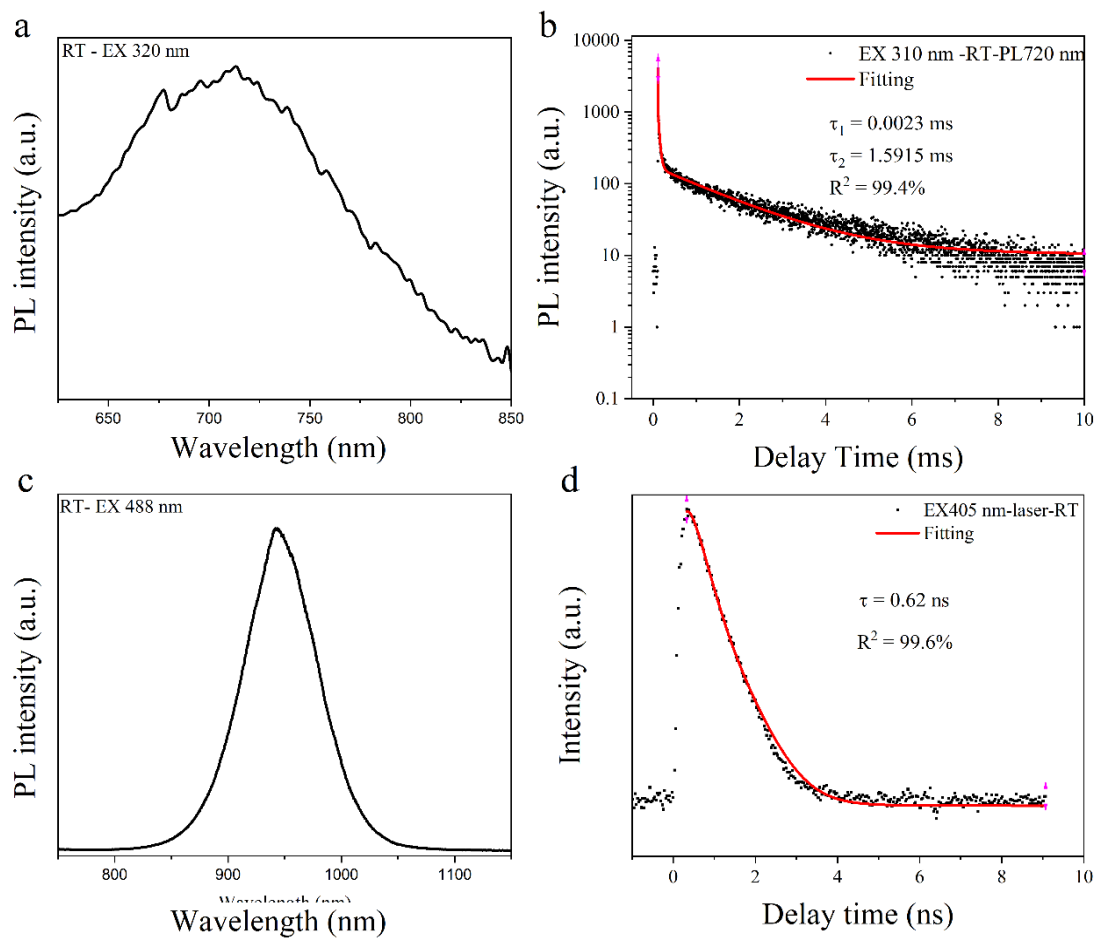

Figure S9 (a) Emission spectrum measured by the Edinburgh FLS1000 spectrometer at room temperature, the emission wavelength is above 720 nm at 320 nm excitation, (b) the measured lifetimes are as follows, $\tau_1$ = 0.0023 ms and $\tau_2$ = 1.5915 ms at 310 nm laser excitation. (c) Measured by WITec confocal microscopy system under 488 nm 20 mW laser excitation, (d) Infrared full spectrum measured by a WITec confocal microscope system under 405 laser (80.0 MHz) excitation with a fitted full spectrum lifetime of 0.62 ns.

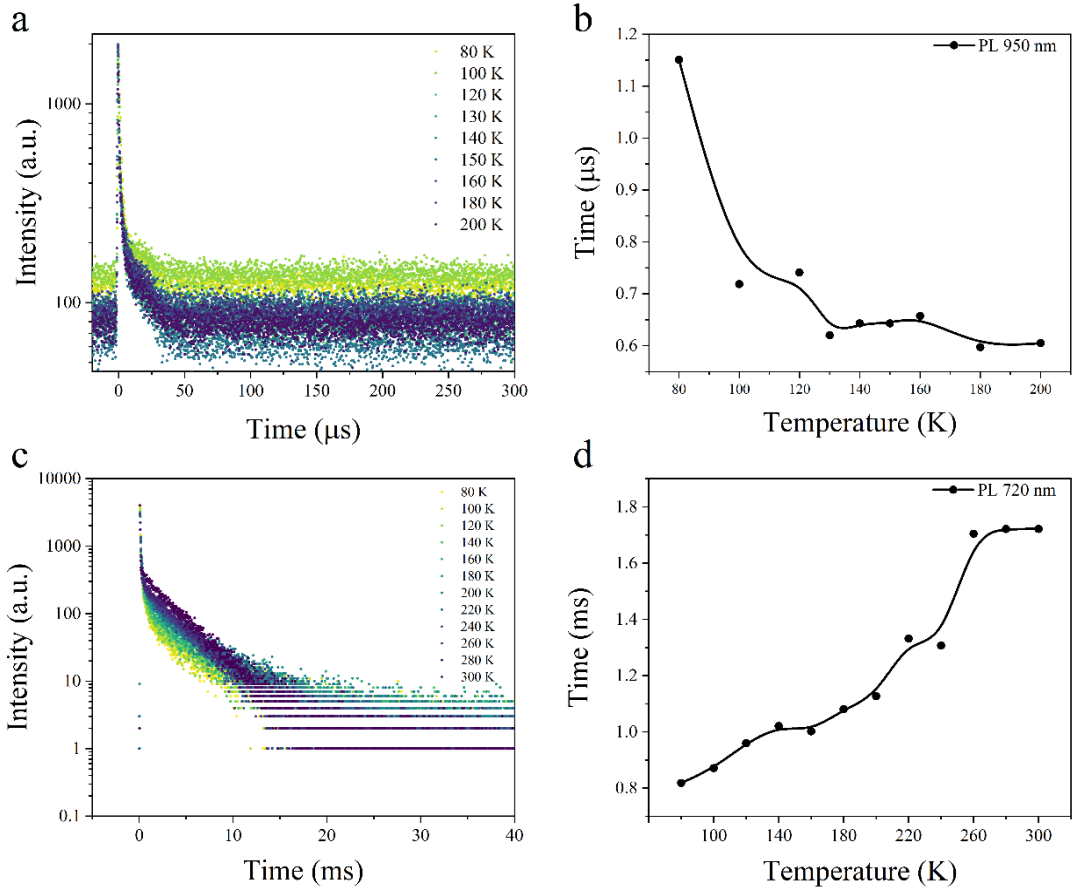

Figure. S10 (a) Lifetime variation of PL at 950 nm in the range of 80-200 K using a HORIBA IR probe (Excitation wavelength 438 nm), (b) is a line graph of the lifetime variation from 80-200 K after a double exponential fit ($y = A_1 e^{-\frac{x}{t_1}} + A_2 e^{-\frac{x}{t_2}} + y_0$). (c) Lifetime change at 80-300 K at $PL_1 = 720$ nm probed using the Edinburgh FLS1000(Excitation wavelength 310 nm laser), (d) the same is a line graph of the lifetime change from 80-300 K after fitting through the double exponential.

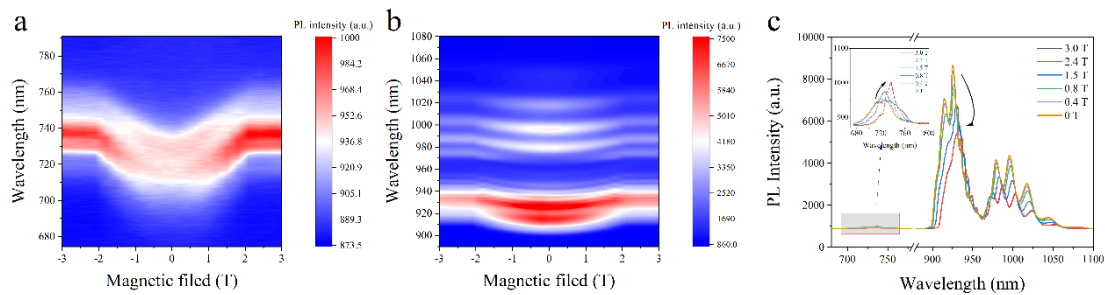

Figure S11 Under different magnetic fields in the 10 K temperature, the luminescence of high-energy $PL_1$ and infrared $PL_2/PL_3$ changes with the variation of positive and negative magnetic field intensities. (a) The change of infrared $PL_2/PL_3$ under different magnetic fields, as the magnetic field increases, the $PL_2/PL_3$ intensity gradually decreases, and the peak wavelength gradually redshifts. (b) The high-energy $PL_1$ increases with the increase of magnetic field strength, and the peak wavelength gradually redshifts. (c) The PL in the range of 680-1100 nm changes with the change of the positive magnetic field.

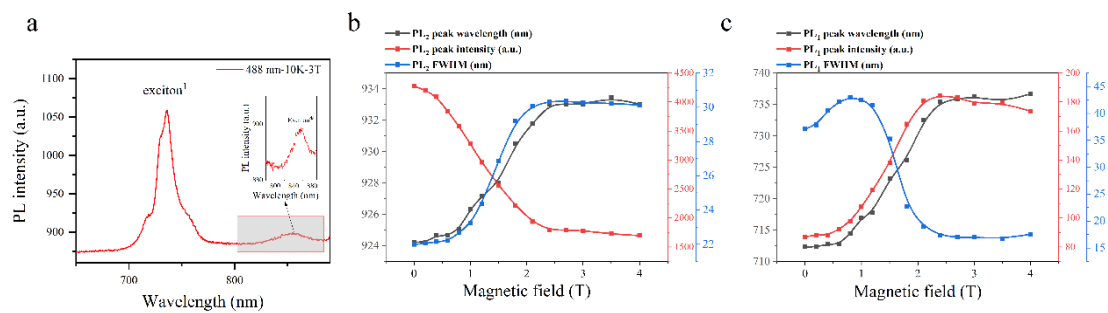

Figure S12 Figure 5 Magneto-PL studies at 10 K: (a) 2D pseudo-color plot of $PL_1$ and dark-state exciton (~850 nm) evolution under magnetic field applied perpendicular to the a-b plane, with 488 nm laser excitation (20 mW). (b) Magnetic field dependence of $PL_1$'s peak wavelength, FWHM, and peak intensity. (c) Magnetic field dependence of $PL_2$'s peak wavelength, FWHM, and peak intensity.

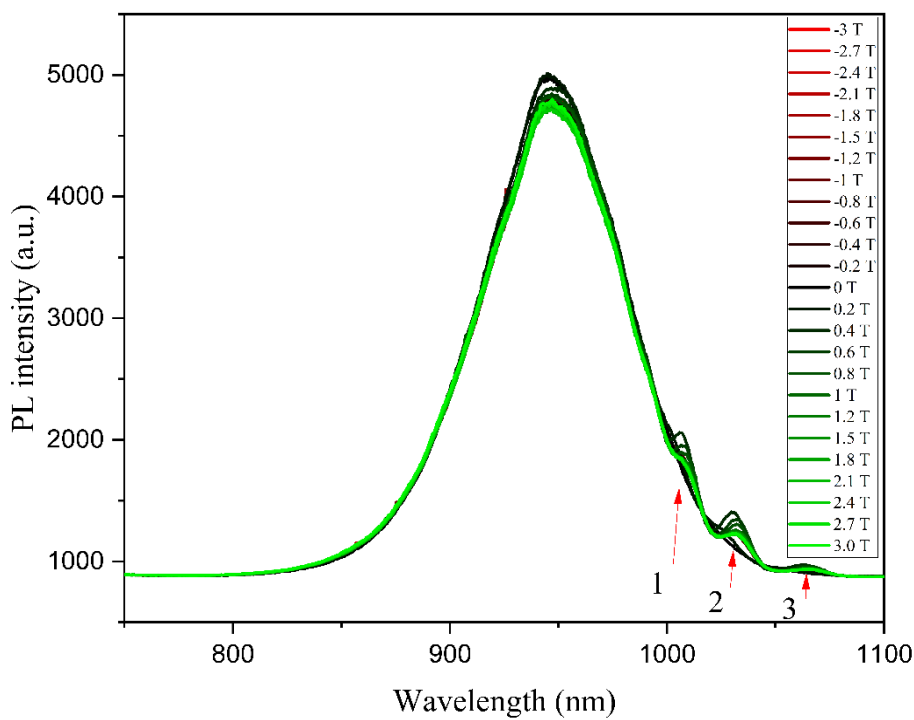

Figure S13 in the magnetic field correlation spectrum at room temperature 20°C, three small peaks appear. The PL peak wavelengths are approximately 1006.48 nm, 1029.79 nm and 1062.63 nm respectively (The difference between them is 223.75 $cm^{-1}$, 299.74 $cm^{-1}$, 523.49 $cm^{-1}$), Moreover, the peak intensity of the $PL_3$ center at 945 nm decreases as the magnetic field intensity increases.

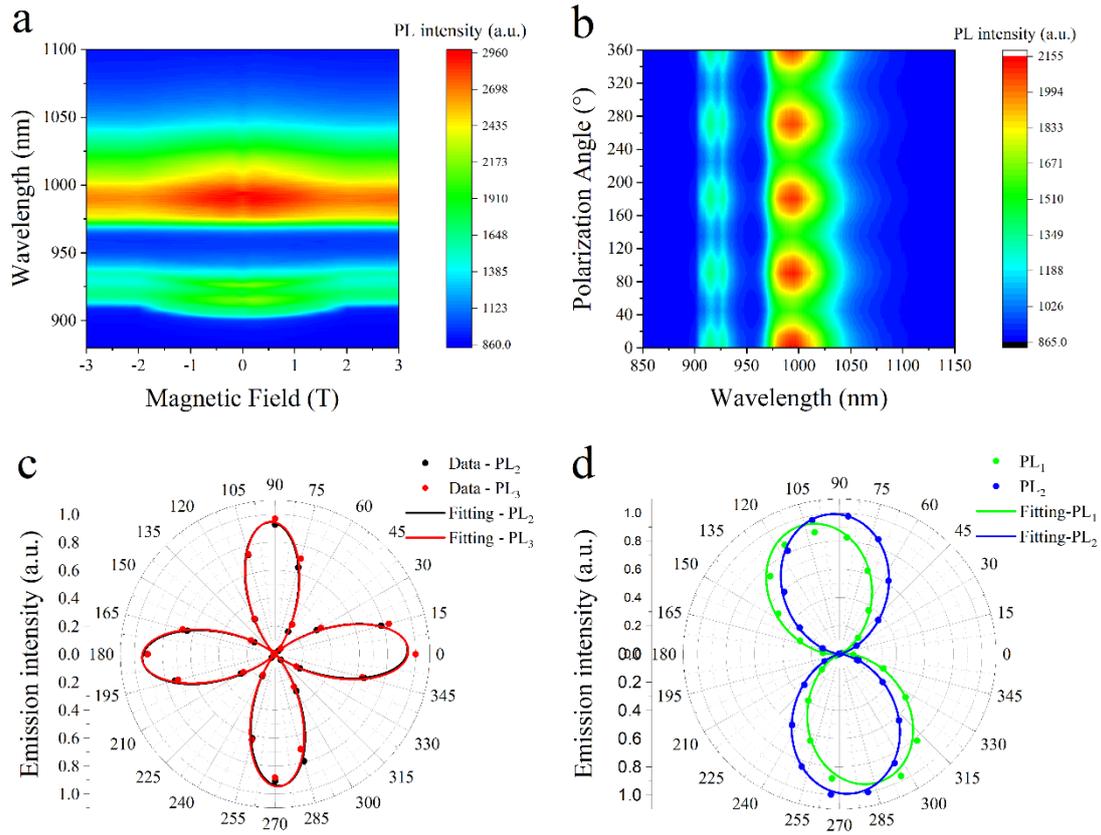

Figure S14 (a) 2D pseudo-color plot showing magnetic field-dependent $PL_2/PL_3$ spectra at 10 K under 400 nm laser excitation (5 mW), demonstrating intensity decrease and red-shift with increasing magnetic field. (b) 2D pseudo-color plot of $PL_2/PL_3$ spectra as a function of half-wave plate rotation angle under 400 nm laser excitation (5 mW). (c) Polar plot of $PL_2/PL_3$ peak intensities versus rotation angle under 400 nm laser excitation (5 mW), with sinusoidal fits showing a 2° difference in polarization direction between $PL_2$ and $PL_3$ (sinusoidal fitting, $R^2$ = 99%). (d) Polarization-resolved measurements of PL peaks at 10 K and 20 mW excitation power (488 nm), using both excitation and detection polarization analysis. The polarization angle difference between $PL_1$ and $PL_2$ is 13.02° (sinusoidal fitting, $R^2$ = 99%).

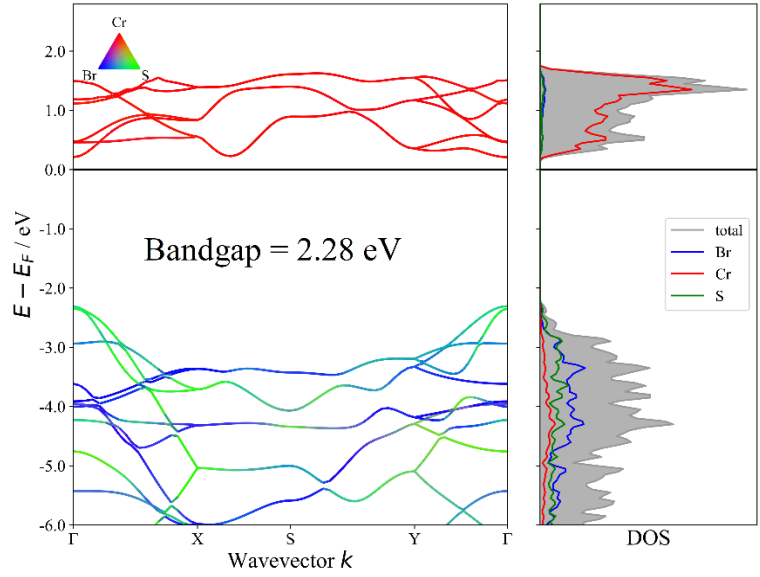

Figure S15 the PBE generalization and the D3G hybrid generalization are used and the DFT+U correction (U = 4 eV) is applied to describe the d-orbital strong correlation effect of $Cr^{3+}$. The calculations cover a single $Cr^{3+}$ ion (6-atom cell) and along the b-axis, using 5 × 5 × 5 k-point grids with a plane-wave truncation energy of 500 eV. The results show a band gap of 2.28 eV for spin reversal along the b-axis in the single cell, with the conduction band contributed mainly by the d-orbitals of $Cr^{3+}$ and the valence band from the $Br^-$ and $S^{2-}$ p orbitals.

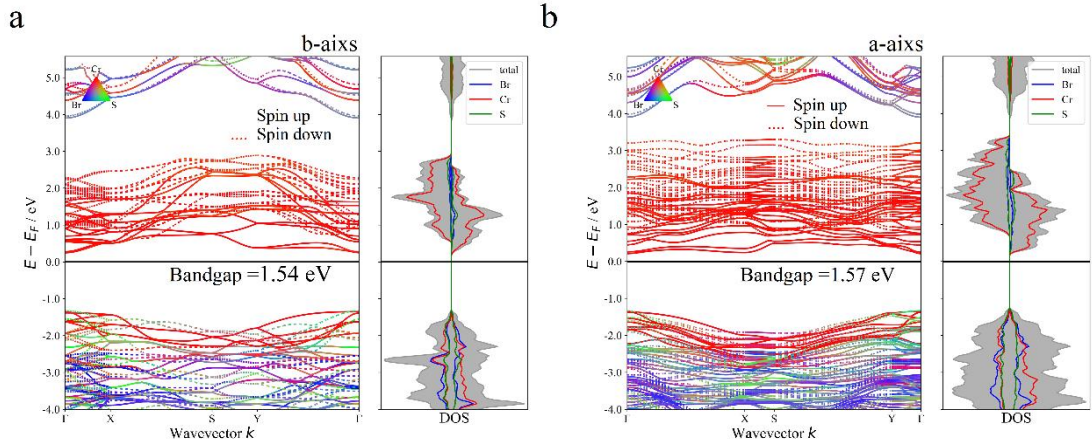

Figure S16 the PBE generalization and the D3G hybrid generalization are used and the DFT+U correction (U = 4 eV) is applied to describe the d-orbital strong correlation effect of $Cr^{3+}$. The calculations cover the three $Cr^{3+}$ spin co-lines of the trimer (24-atom supercell) and along the b- and a-axes, a 17 × 5 × 1 and 6 × 13 × 1 grid of k-points, with a plane-wave truncation energy of 500 eV. The results show that (a) the a-axis spin co-lines have a bandgap value of 1.54 eV at spin-up, and the spin-down bandgap with individual $Cr^{3+}$ ions is close to 2.2 eV. (b) the a-axis spin co-lines have a bandgap value of 1.57 eV for spin up and the band gap with a single $Cr^{3+}$ ion is close to 2.0 eV for spin down.

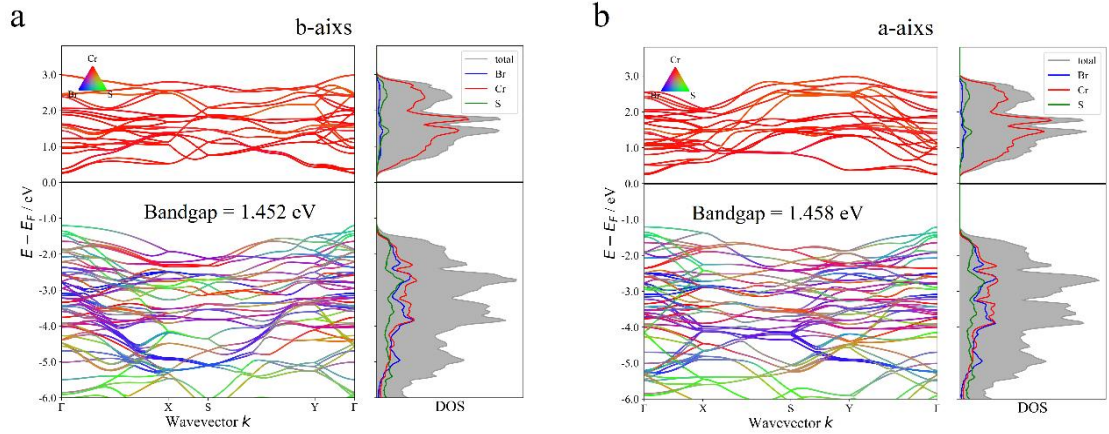

Fig. S17 a hybrid PBE and D3G calculation was used and the DFT+U correction (U = 4 eV) was applied to characterize the d-orbital strong correlation effect of $Cr^{3+}$. Since in our experiments, we observed that $PL_3$ occurs only in multilayers, we hypothesize that it may arise from layer-to-layer aggregation and because of the possibility of a bilayer FM[8], In addition, their $PL_2$ and $PL_3$ polarization directions at 10 K in Fig S12 b-c are essentially the same. The calculations cover two $Cr^{3+}$ spin non-collinear of the upper and lower layers (four under two adjacent mountains per layer) of the tetramer (12-atom supercell) as well as a 17 × 7 × 7 and 9 × 13 × 7 lattice of k-points along the b-axis and the a-axis (with the direction of spin of each $Cr^{3+}$ uniformly oriented toward the b-axis), with plane-wave truncation energies of 500 eV. The results show that (a) the value of the tetramer bandgap facing the b-axis spin noncollinearity is 1.452 eV (b) The value of the tetramer bandgap facing the a-axis spin noncollinearity is 1.458 eV.

# Reference

1    Blöchl, P. E. Projector augmented-wave method. *Physical review B* **50**, 17953 (1994).
2    Kresse, G. & Furthmüller, J. Efficient iterative schemes for ab initio total-energy calculations using a plane-wave basis set. *Physical review B* **54**, 11169 (1996).
3    Perdew, J. P., Burke, K. & Ernzerhof, M. Generalized gradient approximation made simple. *Physical review letters* **77**, 3865 (1996).
4    Datta, B. *et al.* Magnon-mediated exciton-exciton interaction in a van der Waals antiferromagnet. *arXiv preprint arXiv:2409.18501* (2024).
5    Shi, J. *et al.* Giant Magneto-Exciton Coupling in 2D van der Waals CrSBr. *arXiv preprint arXiv:2409.18437* (2024).
6    Esteras, D. L., Rybakov, A., Ruiz, A. M. & Baldoví, J. J. Magnon straintronics in the 2D van der Waals ferromagnet CrSBr from first-principles. *Nano Letters* **22**, 8771-8778 (2022).
7    Rudenko, A. N., Rösner, M. & Katsnelson, M. I. Dielectric tunability of magnetic properties

in orthorhombic ferromagnetic monolayer CrSBr. *npj Computational Materials* **9**, 83 (2023).

8   Sun, Z. *et al.* Resolving and routing the magnetic polymorphs in 2D layered antiferromagnet. *arXiv preprint arXiv:2410.02993* (2024).

## ACKNOWLEDGMENT

The calculation was supported by the high-performance computing platform of Guangxi University. We gratefully acknowledge the support from Hongzhiwei Technology (Shanghai) Co., Ltd., 1599 Xinjinqiao Road, Pudong, Shanghai, China.